\documentclass[a4paper]{jpconf}
\usepackage{graphicx}
\usepackage{xcolor}
\usepackage{amssymb}

\begin{document}
\title{Intrinsic peculiarities of real material realizations of a spin-1/2 kagom\'e lattice}

\author{O Janson$^1$, J Richter$^2$ and H Rosner$^1$}

\address{$^1$ Max-Planck-Institut f\"{u}r Chemische Physik fester
Stoffe, D-01187 Dresden, Germany}
\address{$^2$ Institut f\"{u}r Theoretische Physik, Universit\"{a}t
Magdeburg, D-39016 Magdeburg, Germany}

\ead{rosner@cpfs.mpg.de}

\begin{abstract}
Spin-1/2 magnets with kagom\'e geometry, being for years a generic
object of theoretical investigations, have few real material
realizations. Recently, a DFT-based microscopic model for two such
materials, kapellasite Cu$_3$Zn(OH)$_6$Cl$_2$ and haydeeite
Cu$_3$Mg(OH)$_6$Cl$_2$, was presented $[$Janson O, Richter J and
Rosner H, arXiv:0806.1592$]$. Here, we focus on the intrinsic
properties of real spin-1/2 kagom\'e materials having influence on the
magnetic ground state and the low-temperature excitations. We find that 
the values of exchange integrals are strongly dependent on O---H distance
inside the hydroxyl groups, present in most spin-1/2 kagom\'e
compounds up to date. Besides the original kagom\'{e} model,
considering only the nearest neighbour exchange, we emphasize the
crucial role of the exchange along the diagonals of the kagom\'e
lattice.
\end{abstract}

\begin{figure}[h]
\includegraphics[width=7.5cm]{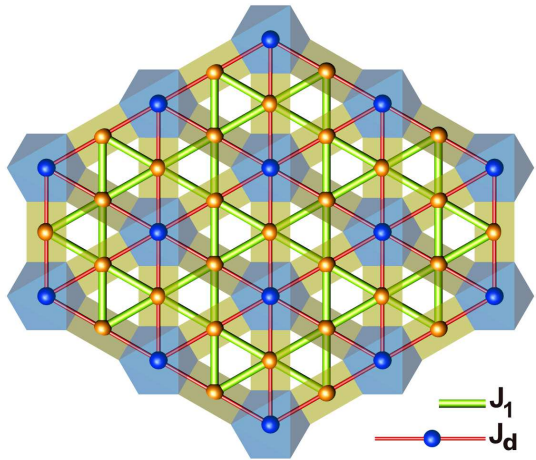}\hspace{2cm}%
\begin{minipage}[b]{5cm}\caption{\label{str}Kagom\'e layers in the
crystal structure of kapellasite (haydeeite). CuO$_4$ plaquettes are
shown in yellow, ZnO$_6$ (MgO$_6$) octahedra are shown in blue. The relevant
exchange paths are highlighted.}
\end{minipage}
\end{figure}

Two-dimensional (2D) magnets with a kagom\'{e} lattice arrangement of
magnetic ions (Figure~\ref{str}) are geometrically frustrated due to
a triangular-like arrangement of nearest neighbours (NN) leading to an
unusual highly degenerate classical ground state
(GS)~\cite{lhuillier01}. Strong quantum fluctuations arising from
spin-1/2 ions can lift the classical degeneracy and drive the system
into a magnetically disordered quantum paramagnetic state, which might
be applied for future quantum computational
applications~\cite{quantum_comp}. Nonetheless, in real materials
couplings to further neighbours~\cite{harris92,our_paper}  and between
the kagom\'{e} layers~\cite{schmal04} are always present and
influence the GS and the thermodynamics.

\begin{figure}[h]
\includegraphics[width=16cm]{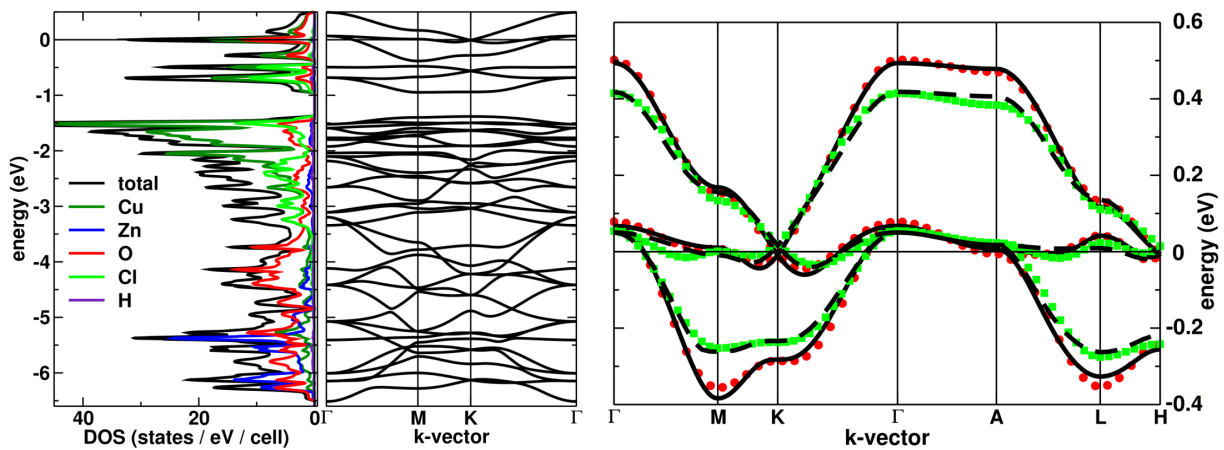}
\caption{\label{dos} Left panel: total and
atom-resolved densities of states of kapellasite
Cu$_3$Zn(OH)$_6$Cl$_2$ (left) and the corresponding band structure
(right). Right panel: band structure of antibonding $dp\sigma$ bands
crossing the Fermi level and the effective one-band tight-binding fit for
different O---H distances (experimental: 0.78~\r{A}, LDA bands ---
dashed lines, TB fit --- green squares; optimized: 1~\r{A}, LDA bands
--- solid lines, TB fit --- red circles).}
\end{figure}

In a recent paper~\cite{our_paper} we have performed DFT calculations
for two new natural isostructural materials with spin-1/2 kagom\'{e}
lattice --- kapellasite Cu$_3$Zn(OH)$_6$Cl$_2$~\cite{kapellasite} and
haydeeite Cu$_3$Mg(OH)$_6$Cl$_2$~\cite{haydeeite,mg_structure}. The local
density approximation (LDA) yields a metallic solution (left panel in
Figure~\ref{dos}) in contrast to the experimentally observed
insulating behaviour due to underestimation of strong on-site
correlations for the Cu$^{2+}$ sites. To account for this deficit, we
carried out an effective one-band tight-binding (TB) fit of the LDA
bands (Figure~\ref{dos}, right panel) and subsequently mapped the
resulting transfer integrals $t_i$ to an extended Hubbard model and to
a Heisenberg model with $J_i=4t_i^2/U_{eff}$.  This mapping provides
an estimate for the antiferromagnetic (AFM) part of the exchange. To
obtain the values for the total exchange, consisting of AFM and
ferromagnetic contributions, we have performed supercell calculations
for various spin arrangements. The correlations were treated in the
LSDA+$U$ approximation.  We have found that besides the standard
kagom\'{e} model based on NN interactions ($J_1$), an additional
coupling along diagonals of kagom\'{e} hexagons ($J_d$) is relevant
(Figure~\ref{str}). Since the absolute values of $J_1$ and $J_d$
change only the overall temperature scale, their ratio
$\alpha{\equiv}J_d/J_1$ is crucial for the GS and low-energy
excitations.

\begin{figure}[h]
\includegraphics[width=16cm]{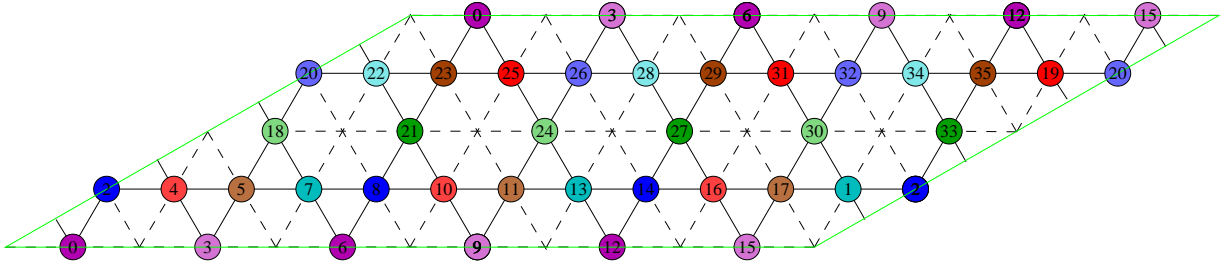}
\caption{\label{36}Finite kagom\'e lattice of $N=36$ sites. The solid
lines represent the NN bonds $J_1$ and the dashed lines the diagonal
bonds $J_d$. The coloured circles indicate the spin orientations of the
twelve-sublattice classical GS relevant for $J_d>0$. Along the chains formed by
diagonal bonds there is an antiparallel (N\'{e}el) spin alignment (e.g. the
spins at the light and dark green circles). On each triangle formed by NN bonds
$J_1$ (e.g. spins on sites 21, 22, 23) there is a $120^{\circ}$ spin
arrangement.  In addition, two antiparallel (N\'{e}el) spin-sublattices are
perpendicular to one other group of two N\'{e}el-like sublattices, i.e.
\textcolor{blue}{ \huge $ \bullet$ } {\large $\perp \;$}
{\color{brown!70!black}\huge $\bullet$
} \quad ; \quad   {\color{green!70!blue} \huge $ \bullet$ } {\large
$\perp \;$} {\color{magenta!70!blue} \huge $\bullet$ }   \quad ; \quad
\textcolor{cyan!70!blue}{ \huge $ \bullet$ } {\large $\perp \;$}
\textcolor{red!90!black}{ \huge $\bullet$ }.}
\end{figure}

\begin{figure}[h]
\includegraphics[width=9.6cm]{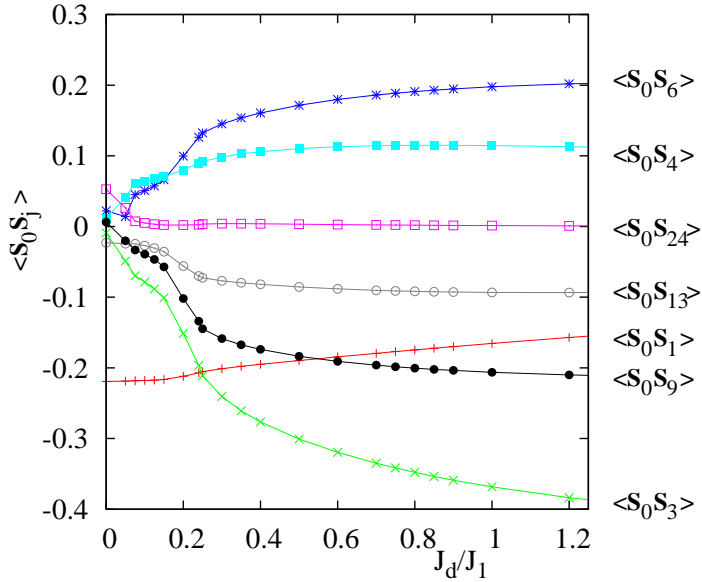}\hspace{0.3cm}%
\begin{minipage}[b]{5.9cm}\caption{\label{sisj}
GS spin-spin correlation $\langle {\bf S}_0{\bf S}_{\bf j}\rangle $ in
dependence on $\alpha=J_d/J_1$ for the finite lattice of $N=36$ sites
shown in Figure~\ref{36}. The site indices $j=1,3,4,6,9,13,24$
correspond to those of Figure~\ref{36}.
\hspace{3cm} $ $
\hspace{3cm} $ $
\hspace{3cm} $ $
\hspace{3cm} $ $}
\end{minipage}
\end{figure}

One focus of this paper is to emphasize the influence of side groups
(which are commonly neglected in model physics) on different exchange
paths. A key issue of theoretical investigations of real materials is
the construction of a relevant model, which describes experimental
data and has predictive power. For cuprates, that are insulators with
a $3d^9$ Cu$^{2+}$ configuration, the extension from the initial
Heisenberg idea to consider only spin degrees of freedom (Heisenberg
model) to a multiband Hubbard model with a separate treatment of Cu
$3d$ and O $2p$ orbitals became computationally feasible only recently. On
the other hand, the famous Goodenough-Kanamori-Anderson (GKA) rules formulated
quite early on an empirical basis provide a simple intuitive picture
for the exchange interactions based on geometrical quantities. Due to
the simplicity of the GKA rules, they are violated in many cases. A natural
way to search for a possible origin of these phenomena is to improve
the model by including ligand fields effect. In general, such
models are very difficult to evaluate. Even in case of simple systems
like CuGeO$_3$ the models are quite complicated and contain many
parameters~\cite{CuGeO3_side_groups}.

Here, we present the results of
a DFT-based modeling for Cu$_3$Zn$[$Mg$]$(OH)$_6$Cl$_2$. In these
compounds, the leading exchange interactions are strongly dependent on
the position of H atoms --- the simplest possible side groups.

For haydeeite, the experimentally defined H position yields an
unusually short O---H distance ($d$) of 0.78~\r{A} (for kapellasite the H
position has not been reported). Therefore, in LDA calculations we
relaxed the H position which resulted in an optimized
$d\approx1$~\r{A} for both systems. In order to elucidate the
influence of $d$ on the exchange integrals, we have
studied the range $d=1.0\pm0.3$~\r{A} which covers the experimental
errors (that are large due to weak x-ray scattering by H atoms). The
details of our calculations are described in~\cite{our_paper}. We find
that $d$ has a strong influence on the NN exchange $J_1$, while $J_d$
is almost unaffected.  Obviously, this modifies their ratio $\alpha$ and
consequently the physical properties. The underlying reason is the
shift of oxygen states down in energy caused by the shortening
of an O---H bond. In consequence, Cu---O hybridization decreases.
Therefore, for a fixed Cu---O geometry, a decreased $d$ reduces the AFM
contribution to the total exchange. This is supported by a strongly
reduced bandwidth (which is related via $t_i^2{\sim}J_i^{AFM}$ to the AFM exchange) for
a reduced $d$ (Figure~\ref{dos}, right panel) and fits well
the results of supercell calculations. There, we find a quasi-linear $J_1(d)$
dependence with a positive slope for both compounds.  $J_1$ is 2.5
times larger for $d=1$~\r{A} than for $d=0.8$~\r{A} in kapellasite
(2.5~meV versus 1~meV), while it changes the sign in haydeeite
($0.7$~meV versus $-0.2$~meV).

The next step towards a physically relevant picture of real materials
is to evaluate the influence of $\alpha$ onto the GS and the low-lying
excitations. Here, we use a $J_1$-$J_d$ spin-1/2 Heisenberg model on
the kagom\'e lattice ($J_1$-$J_d$ model).  The classical GS of the pure
kagom\'e system ($J_d=0$) is known to be highly degenerate.  The additional
diagonal bond $J_d$ reduces this degeneracy drastically and selects
non-coplanar classical GS's with twelve magnetic sublattices (see
Figure~\ref{36}) among the huge number of classical GS's existing for $J_d=0$.
For the quantum spin-1/2 model we have calculated the GS by exact
diagonalization for the finite lattice shown in Figure~\ref{36}. 

Obviously, the quantum GS spin-spin correlations are drastically
affected when including $J_d$. While for \textsl{J}$_d=0$ all spin
correlations except the NN correlation function  $\langle {\bf
S}_0{\bf S}_{\bf 1}\rangle $ are close to zero we find a well
pronounced short-range order for $\alpha\gtrsim0.25$, that corresponds
to the classical magnetic structure. For $\alpha\gtrsim0.55$
the strongest spin correlations are $\langle {\bf S}_0{\bf S}_{\bf
3}\rangle $, $\langle {\bf S}_0{\bf S}_{\bf 6}\rangle $, $\langle {\bf
S}_0{\bf S}_{\bf 9}\rangle $, and they  belong to the chains formed by
$J_d$ bonds. The NN correlation function $\langle {\bf S}_0{\bf
S}_{\bf 1}\rangle $ monotonously decreases with $J_d$. The
correlation function $\langle {\bf S}_0{\bf S}_{\bf 24}\rangle $
belongs to two sites located on sublattices being perpendicular in the
classical GS, and it is almost zero for $\alpha\gtrsim0.55$. This
leads to the conclusion that even in the quantum model the GS might
have a non-coplanar magnetic structure giving rise to enhanced chiral
correlations. 

To summarize, we have shown the relevance of the ligand field for the
low-energy physics by example of two spin-1/2 cuprates with a kagom\'{e}
lattice geometry. We found that there are two relevant exchange
integrals in both materials. The ratio $\alpha=J_d/J_1$ is strongly
dependent on the O---H distance and has a drastic impact on the
physical properties: the quantum GS of a corresponding spin-1/2
Heisenberg $J_1$-$J_d$ antiferromagnet on the kagom\'e lattice
exhibits strong magnetic correlations along $J_d$ bonds for
$\alpha>0.5$, whereas the other correlations remain weak. Therefore,
the low-energy excitations might be $S=1/2$ spinons causing an
effectively one-dimensional low-temperature physics as has been
discussed previously for other 2D models such as the crossed-chain
model~\cite{Starykh_PRL_02} as well as for the anisotropic triangular
lattice~\cite{Coldea_PRL_01}. As the physical properties are related
to $\alpha$ which is strongly dependent on the O---H distance, the H
position should be reinvestigated experimentally with better accuracy. 
Generally, a precise determination of the position of side group
atoms is crucial for an accurate model derivation. We suggest that the
dependence of exchange integrals on side group positions can be used
to tune a magnetic ground state under changed conditions like external
pressure.

\section*{References}

\end{document}